\documentclass[a4paper,11pt]{article}
\usepackage{pos}

\title{Type IIB S-folds: flat deformations, holography and stability}

\author*[a,b]{Adolfo Guarino}
\author[c,a]{Colin Sterckx}

\affiliation[a]{Departamento de F\'isica, Universidad de Oviedo,\\
Avda. Federico Garc\'ia Lorca 18, 33007 Oviedo, Spain.}

\affiliation[b]{Instituto Universitario de Ciencias y Tecnolog\'ias Espaciales de Asturias (ICTEA) \\
Calle de la Independencia 13, 33004 Oviedo, Spain.}

\affiliation[c]{Universit\'e Libre de Bruxelles (ULB) and International Solvay Institutes,\\
Service  de Physique Th\'eorique et Math\'ematique, \\
Campus de la Plaine, CP 231, B-1050, Brussels, Belgium.}

\emailAdd{adolfo.guarino@uniovi.es}
\emailAdd{colin.sterckx@ulb.ac.be}

\abstract{We review recent progress in the study of S-folds in light of the gauge/gravity duality and \mbox{the AdS swampland conjecture}. S-folds correspond to non-geometric backgrounds of type IIB \mbox{supergravity} of the form $\,\textrm{AdS}_4 \, \times \, \textrm{S}^1 \, \times \, \mathcal{M}\,$ that involve a non-trivial $\,\textrm{SL}(2,\,\mathbb{Z})\,$ (S-duality) \mbox{monodromy} for the type IIB fields when moving around the $\,\textrm{S}^1$. We present four such solutions with $\,\mathcal{M}=\textrm{S}^{5}\,$ that preserve $\,\mathcal{N}=4,2,1,0\,$ supersymmetries. Via the AdS/CFT correspondence, these solutions are conjectured to describe new strongly coupled three-dimensional CFT's on a localised interface of SYM. We discuss the existence of flat deformations in the gravity side dual to marginal deformations of the conjectured S-fold CFT's. From a geometrical perspective, the flat deformations induce a monodromy $\,h\,$ on $\,\mathcal{M}\,$ and replace $\,\textrm{S}^1 \,\times\, \mathcal{M}\,$ by the so-called mapping torus $\,T(\mathcal{M})_h$. Interestingly, the flat deformations provide a controlled mechanism of supersymmetry breaking for $\,\mathcal{N} \ge 2\,$ S-folds. We present a class of such non-supersymmetric S-folds obtained by flat-deforming the $\,\mathcal{N}=4\,$ S-fold and discuss their (non-)perturbative stability.}

\FullConference{%
  Corfu Summer Institute 2021 "School and Workshops on Elementary Particle Physics and Gravity"\\
  29 August - 9 October 2021\\
  Corfu, Greece
}

\begin{document}
\maketitle

\section{Introduction}

String Theory can accommodate non-geometric vacuum configurations for which a generalised notion of geometry, going well beyond the fundamentals of Riemannian geometry, is required in order to understand how geometrical and gauge aspects get intertwined with one another. This lies at the core of the so-called \textit{string dualities} which establish the physical equivalence between seemly different string backgrounds \cite{Hull:1994ys}. A distinctive feature of such \textit{generalised geometries} is that the tangent space of the internal manifold gets extended to include transition functions between different patches that involve not only diffeomorphisms but also gauge transformations for the various $p$-form fields in the low energy supergravity description of the stringy background \cite{Coimbra:2011nw,Coimbra:2012af}. Even more strikingly, the set of available transition functions includes transformations in the duality group (typically an exceptional Lie group in the case of maximally supersymmetric theories \cite{Cremmer:1997ct}) which in turn go well beyond internal diffeomorphisms and gauge transformation, thus generating a so-called \textit{non-geometric} background (see \cite{Plauschinn:2018wbo} for a recent review).

Among the non-geometric string backgrounds, a particular class of type IIB backgrounds, dubbed S-folds in \cite{Hull:2004in}, have recently received special attention. These type IIB backgrounds involve, as transition functions in the internal geometry, elements of the non-perturbative S-duality group $\,\textrm{SL}(2,\,\mathbb{Z})\,$ of type IIB string theory \cite{Hull:2003kr,Hull:2004in}. In its simplest incarnation, an S-fold background is described by a ten-dimensional geometry of the form $\,\textrm{AdS}_{4} \times \textrm{S}^{1} \times \textrm{S}^5\,$ in which the type IIB fields transforming non-trivially under S-duality undergo a non-trivial $\,\textrm{SL}(2,\,\mathbb{Z})\,$ monodromy when making a loop around the $\,\textrm{S}^{1}$ \cite{Inverso:2016eet}. As a consequence of the non-trivial $\,\textrm{SL}(2,\,\mathbb{Z})\,$ monodromy, the resulting type IIB background is locally well-defined but cannot be extended to a globally well-defined one. 

One of the most challenging goals of the S-fold program is to identify and characterise the corresponding field theory duals in light of the AdS/CFT correspondence. Featuring an AdS$_{4}$ factor in the ten-dimensional geometry, the type IIB S-fold backgrounds are conjectured to be dual to a new class of strongly-coupled three-dimensional CFT's dubbed S-fold CFT's \cite{Assel:2018vtq}. In the gravity side, a method to construct type IIB S-fold backgrounds was presented in \cite{Bobev:2020fon}. More concretely, S-folds can be systematically generated upon taking a special limit on Janus solutions in the effective five-dimensional SO(6)-gauged supergravity description of type IIB compactified on $\,\textrm{S}^{5}$. Due to the holographic correspondence between (five-dimensional) Janus solutions and interfaces in $\,\mathcal{N}=4\,$ super-Yang–Mills (SYM) theory \cite{Bak:2003jk,Clark:2004sb,DHoker:2006qeo}, it becomes natural to identify the S-fold CFT's with new strongly coupled three-dimensional CFT's that are localised on an interface of $\,\textrm{SYM}$. For the original $\,\mathcal{N}=4\,$ S-fold with SO(4) symmetry put forward in \cite{Inverso:2016eet}, a dual S-fold CFT was proposed in \cite{Assel:2018vtq}. Such an S-fold CFT would appear as the effective IR description of a three-dimensional $\,\textrm{T[U($N$)]}\,$ theory \cite{Gaiotto:2008ak} where the diagonal subgroup of the $\,\textrm{U}(N)^2\,$ flavour group is gauged using an $\mathcal{N} = 4\,$ vector multiplet and where a Chern--Simons term at level $\,k\,$ is also activated. It is precisely the Chern--Simons level $\,k\,$ the one responsible for the hyperbolic $\,J_k \in \textrm{SL}(2,\,\mathbb{Z})\,$ monodromy along the $\,\textrm{S}^1\,$ characterising the type IIB S-fold background in the gravity side.

In this proceeding we describe some recent progress in the S-fold program. More concretely, we discuss various examples of S-fold backgrounds constructed in \cite{Guarino:2019oct,Guarino:2020gfe} in addition to the original $\,\mathcal{N}=4 \,\&\, \textrm{SO}(4)\,$ symmetric S-fold of \cite{Inverso:2016eet}. In all the examples, the $\,\textrm{SL}(2,\,\mathbb{Z})\,$ monodromy is of hyperbolic type, namely, $\,J_k =  - \mathcal{S} \mathcal{T}^{k}\,$ in terms of the standard inversion $\,\mathcal{S}\,$ and unit translation $\,\mathcal{T}\,$ matrices generating $\,\textrm{SL}(2,\,\mathbb{Z})$. Instead of pushing further the five-dimensional approach that uses Janus solutions as a seed to generate S-fold backgrounds, we will adopt an alternative four-dimensional approach. One may wonder whether this is at all possible as the $\textrm{AdS}_{4} \times \textrm{S}^{1}\,$ piece of the S-fold geometry, combined with the non-trivial dependence of the type IIB fields on the $\,\textrm{S}^{1}\,$ due to the $\,\textrm{SL}(2,\,\mathbb{Z})\,$ monodromy, seem to render the problem intrinsically five-dimensional. However, employing techniques of Exceptional Field Theory \cite{Hohm:2013pua,Hohm:2013uia,Hohm:2014qga}, it was shown in \cite{Inverso:2016eet} that type IIB supergravity admits a consistent generalised Scherk--Schwarz reduction on $\,\textrm{S}^{1} \times \textrm{S}^{5}\,$ that implements the S-duality twist responsible for the $\,\textrm{SL}(2,\,\mathbb{Z})\,$ monodromy. The outcome of such a generalised Scherk--Schwarz reduction is an effective four-dimensional maximal gauged supergravity with a gauging $\,[\textrm{SO}(1,1) \times \textrm{SO}(6)] \ltimes \mathbb{R}^{12}\,$ of dyonic type \cite{Dall'Agata:2014ita}.

One of the main advantages of working within the realm of four-dimensional supergravity is that any complicated S-fold background in ten-dimensions simply corresponds to a maximally symmetric $\,\textrm{AdS}_{4}\,$ solution in the four-dimensional effective theory. This renders the problem purely algebraic in four dimensions and finding S-fold solutions amounts to extremise the scalar potential of the effective four-dimensional supergravity. Once an $\,\textrm{AdS}_{4}\,$ vacuum is found in four dimensions, its ten-dimensional type IIB uplift can be straightforwardly worked out using the generalised Scherk--Schwarz Ansatz following the results in \cite{Inverso:2016eet}. A second advantage of working within the four-dimensional effective theory is that one gets a better control on the set of possible S-fold deformations. Lastly, assessing the perturbative (in)stability of the S-fold backgrounds becomes also easier from a four-dimensional perspective due to the Kaluza-Klein spectrometry techniques recently developed in \cite{Malek:2019eaz}.

Moving to the field theory side, S-fold CFT's are expected to feature a conformal manifold of marginal deformations including, at least, those marginal deformations inherited from their interface counterparts \cite{DHoker:2006qeo}. Such marginal deformations in the field theory side should appear as flat directions of the supergravity scalar potential by virtue of the $\textrm{AdS}_{4}/\textrm{CFT}_{3}$ correspondence. In this proceeding we focus on a particular class of such \textit{flat deformations} of the supergravity scalar potential and provide an interpretation thereof in five and ten dimensions. When oxidised to five dimensions, this class of flat deformations corresponds to turning on one-form (Wilson lines) deformations in a putative five-dimensional background \cite{Guarino:2021kyp}. From a ten-dimensional perspective, the flat deformations generically cause a global breaking of isometries in the internal manifold from the original isometry group of the undeformed solution down to its Cartan subgroup. More concretely, as shown in \cite{Guarino:2021kyp} (see also \cite{Giambrone:2021zvp}), the flat deformations induce a geometrical monodromy $\,h\,$ on the internal $\,\textrm{S}^5\,$ when moving around the $\,\textrm{S}^1$. This geometrical monodromy $\,h\,$ -- to be distinguished from the $\,\textrm{SL}(2,\,\mathbb{Z})\,$ S-duality monodromy $\,J_{k}\,$ -- induces various patterns of symmetry breaking as classified by the mapping torus $\,T_{h}(\textrm{S}^{5})\,$ \cite{Guarino:2021kyp}.

As a particular example of the above story, we concentrate on the original $\,\mathcal{N}=4 \,\&\, \textrm{SO}(4)\,$ symmetric S-fold of \cite{Inverso:2016eet} and establish the existence of two such flat deformations. Turning on the deformations generically breaks the original $\,\textrm{SO}(4)\,$ symmetry down to its $\,\textrm{U}(1)^2\,$ Cartan subgroup. In addition, it also breaks \textit{all} of the original supersymmetries. However, as shown recently in \cite{Giambrone:2021zvp}, this does \textit{not} trigger any perturbative instability in the ten-dimensional S-fold background so it remains perturbatively stable. We will conclude this proceeding by briefly commenting on further implications of our results in light of the AdS swampland conjecture \cite{Ooguri:2016pdq}.

\section{S-fold solutions of type IIB supergravity}
\label{sec:S-fold_solutions}

S-folds are solutions of the source-less equations of motion and Bianchi identities of type IIB supergravity in which the geometry is of the form $\,\textrm{AdS}_4 \times \textrm{S}^1 \times \mathcal{M}$ (we will focus on $\,\mathcal{M}=\textrm{S}^5\,$). Their main distinctive feature is that the non-trivial dependence on the coordinate $\,\eta\,$ along $\,\textrm{S}^{1}\,$ is totally encoded in an $\,\textrm{SL}(2,\mathbb{R})\,$ S-duality twist
\begin{equation}
\label{A-twist}
A^\alpha{}_\beta(\eta) = \begin{pmatrix} e^{-\eta} & 0\\ 0 & e^{\eta} \end{pmatrix} \ .
\end{equation}
The ten-dimensional metric is given by
\begin{equation}
\label{metric_10D_gen}
ds_{10}^2 = \Delta^{-1}  \left[\, \tfrac{1}{2} ds_{\textrm{AdS}_4}^2+ds_{6}^2\right] \ ,
\end{equation}
with $\,\Delta\,$ being the warping factor. A shift along $\,\eta\,$ must be an isometry of the metric (\ref{metric_10D_gen}) as the Einstein-frame metric field is a singlet under S-duality. By the same token, the four-form potential $\,C_{4}\,$ cannot depend on $\,\eta\,$ either. However, the two-form potentials $\,\mathbb{B}^{\alpha}=(B_{2},C_{2})\,$ and the axion-dilaton matrix
\begin{equation}
\label{axion-dilaton_10D}
m_{\alpha\beta} =  \begin{pmatrix} e^{-\Phi} + e^{\Phi} \, C_{0}^2 & - e^{\Phi} \, C_{0} \\  - e^{\Phi} \, C_{0} & e^{\Phi}  \end{pmatrix} \ ,
\end{equation}
have a non-trivial dependence on $\,\eta\,$ as they transform under S-duality. The entire dependence of these type IIB fields on $\,\eta\,$ is encoded in the S-duality twist matrix (\ref{A-twist}). More concretely one has
\begin{equation}
\label{twoForm_10D_gen}
\mathbb{B}^\alpha = A^\alpha{}_\beta \, \mathfrak{b}^\beta
\hspace{8mm} \textrm{ and } \hspace{8mm}
m_{\alpha\beta} = (A^{-t})_{\alpha}{}^{\gamma} \, \mathfrak{m}_{\gamma\delta}  (A^{-1})^{\delta}{}_{\beta} \ ,
\end{equation}
with $\,A^{-t} \equiv (A^{-1})^{t}$.

The coordinate $\,\eta\,$ can be taken to be periodic with period $\,T$. However, due to the $\,\textrm{SL}(2,\,\mathbb{R})\,$ twist in (\ref{A-twist}), there is a non-trivial monodromy
\begin{equation}
\label{monodromy_S1}
\mathfrak{M}_{\textrm{S}^{1}} = A^{-1}(\eta) \, A(\eta + T) =  \begin{pmatrix} e^{-T} & 0\\ 0 & e^{T} \end{pmatrix} \ ,
\end{equation}
of hyperbolic type when making a loop around the $\,\textrm{S}^{1}\,$. This renders the S-fold backgrounds locally geometric but globally non-geometric. The monodromy in (\ref{monodromy_S1}) can be brought into a generic hyperbolic monodromy $\,J_k \in \textrm{SL}(2,\,\mathbb{Z})\,$ of the form
\begin{equation}
J_k = 
\begin{pmatrix}
    k & 1\\
    -1 & 0 
\end{pmatrix} = - \mathcal{S} \, \mathcal{T}^k 
\hspace{5mm} \textrm{with} \hspace{5mm} k \in \mathbb{N} 
\hspace{5mm} \textrm{and} \hspace{5mm}
k \geq 3 \ ,
\end{equation}
provided the period $\,T\,$ becomes $k$-dependent and given by
\begin{equation}
T(k) = \log(k+\sqrt{k^2-4}) - \log(2) \ .
\end{equation}
The quotient by the above type of dualities prevents the S-fold solutions from entering the non-perturbative regime \cite{Assel:2018vtq}. Since the string coupling is given by $\,g_{s} = e^\Phi \propto e^{-2\eta} \, \mathfrak{m}_{22}\,$, there always exists a frame in which $\,g_{s} \ll 1$. This ensures that one can safely work in a perturbative regime of type IIB string theory.

\subsection{$\mathcal{N}=4$ and $\,\textrm{SO}(4)\,$ symmetric S-fold}\label{subsection:N4sol}
\label{sec:N=4_solution}

This S-fold solution appeared originally in \cite{Inverso:2016eet} and has later been reinterpreted in \cite{Giambrone:2021wsm}. Taking advantage of the $\,\textrm{SO}(4) \sim \textrm{SO}(3)_1 \times \textrm{SO}(3)_2\,$ symmetry of the solution, which is identified with the R-symmetry group of the dual $\,\mathcal{N}=4\,$ S-fold CFT, it proves convenient to describe the five-sphere $\,\textrm{S}^{5}\,$ as a product of two two-spheres $\,\textrm{S}^{2}_{i=1,2}\,$ fibered over an interval $\,I$. Each of these two-spheres displays an $\,\textrm{SO}(3)\,$ symmetry. We choose coordinates such that the two-spheres are parameterised by standard polar and azimutal angles $\,(\theta_i,\,\varphi_i)\,$ with $\,\theta_i \in [0,\pi]\,$ and $\,\varphi_i \in [0,2\pi]$, and the interval is parameterised by an angle $\,\alpha \in [0,\,\tfrac{\pi}{2}]$. Then, the internal $\,\textrm{S}^{1} \times \textrm{S}^{5}\,$ metric in (\ref{metric_10D_gen}) is given by
\begin{equation}
\label{metric_10D_solu_N4}
ds_{6}^2 = d\eta^2 + d\alpha^2  + \dfrac{\cos^2\alpha}{2+\cos(2\alpha)} ds_{\textrm{S}^2_1}^2  + \dfrac{\sin^2\alpha}{2-\cos(2\alpha)} ds_{\textrm{S}^2_2}^2  \ ,
\end{equation}
with
\begin{equation}
\label{spheres_S2}
ds_{\textrm{S}^2_i}^2  = d\theta_{i}^2 + \sin^2\theta_{i} \, {d\varphi_{i}}^2 \hspace{8mm} , \hspace{8mm}\textrm{vol}_i = \sin\theta_i\, d\theta_i \wedge d\varphi_i \ ,
\end{equation}
and the non-singular warping factor reads
\begin{equation}
\Delta^{-4} = 4 - \cos^2(2\alpha) \ . 
\end{equation}
The $\,\eta$-independent two-form potentials in (\ref{twoForm_10D_gen}) read
\begin{equation}
\label{B2-C2_solu_N4}
\mathfrak{b}_{1} = - 2 \sqrt{2} \, \dfrac{\cos^3\alpha}{2 + \cos(2\alpha)} \, \textrm{vol}_{1} \hspace{5mm},\hspace{5mm}
\mathfrak{b}_2 =- 2 \sqrt{2} \, \dfrac{\sin^3\alpha}{2 - \cos(2\alpha)} \, \textrm{vol}_{2} \ ,
\end{equation}
whereas the axion-dilaton matrix is
\begin{equation}
\mathfrak{m}_{\alpha\beta} = 
\begin{pmatrix} \sqrt{\frac{2 + \cos(2\alpha)}{2-\cos(2\alpha)}} & 0 \\
0 & 
\sqrt{\frac{2 - \cos(2\alpha)}{2 + \cos(2\alpha)}}
\end{pmatrix}  \ .
\end{equation}
The self-dual five-form is given by
\begin{equation}
\label{F5_solu_N4}
\widetilde{F}_5 = \Delta^4 \, \sin^2(2\alpha) \, (1+\star) \left[ - \tfrac{3}{2} \, \textrm{vol}_5 + \sin(2\alpha) \, d\eta \wedge \text{vol}_{1}\wedge\text{vol}_{2} \right] \ ,
\end{equation}
where $\,\textrm{vol}_5=d\alpha \wedge \textrm{vol}_{1} \wedge \textrm{vol}_{2}$. Lastly, the $\,\textrm{AdS}_{4}\,$ radius in the external part of the metric (\ref{metric_10D_gen}) is set to $\,L^2=1$.

\subsection{$\mathcal{N}=2$ and $\,\textrm{SU}(2)\times\textrm{U}(1) \,$ symmetric S-fold}
\label{sec:N=2_solution}

This S-fold was put forward in \cite{Guarino:2020gfe}. The $\,\textrm{SU}(2) \times \textrm{U}(1)\,$ symmetry of the solution becomes manifest when describing the five-sphere $\,\textrm{S}^{5}\,$ as a three-sphere $\,\textrm{S}^{3}\,$ fibered over a two-sphere $\,\textrm{S}^{2}$. We choose standard polar $\,\theta \in [0,\,\pi]\,$ and azimutal $\,\phi \in [0,\,2\pi]\,$ angles to describe the two-sphere, as well as three angular coordinates $\,\alpha \in [0,\,2\pi]$, $\,\beta \in [0,\,\pi]\,$ and $\,\gamma \in[0,\,4\pi]\,$ to describe the three-sphere. The latter is better described using a set $\,\sigma_{1,2,3}\,$ of $\,\text{SU}(2)\,$ left-invariant one-forms
\begin{equation}
\label{SU2-inv-forms}
\begin{array}{lll}
\sigma_1 &=& \frac{1}{2}\left(- \sin\alpha\, d\beta + \cos \alpha \sin \beta\, d\gamma\right) \ , \\[2mm]
\sigma_2 &=& \frac{1}{2} \left(\cos\alpha\, d\beta +\sin\alpha \sin\beta\, d\gamma\right) \ , \\[2mm]
\sigma_3 &=& \frac{1}{2} \left(d\alpha + \cos\beta\, d\gamma \right) \ .
\end{array}
\end{equation}
In terms of the above one-forms, the internal $\,\textrm{S}^{1} \times \textrm{S}^{5}\,$ metric in (\ref{metric_10D_gen}) is given by
\begin{equation}
\label{metric_10D_solu_N2}
ds_{6}^2 = \tfrac{1}{2} \left( \, d\eta^2 + ds_{\textrm{S}^2}^2  + \cos^2\theta\left[ 8 \, \Delta^4 \, \left( \sigma_1^2 + \sigma_2^2 \right) +  \sigma_3^2 \right] \, \right) \ , 
\end{equation}
with $\,ds_{\textrm{S}^2}^2 = d\theta^2 + \sin^2\theta \, d\phi^2$, and the warping factor reads
\begin{equation}
\Delta^{-4} = 6-2\cos(2\theta) \ .
\end{equation}
The $\,\textrm{U}(1)\,$ factor of the S-fold symmetry group is then realised as rotations in the $\,(\sigma_{1},\sigma_{2})$-plane generated by translations along the coordinate $\,\alpha$. This is identified with the R-symmetry group in the dual $\,\mathcal{N}=2\,$ S-fold CFT.

The $\,\eta$-independent two-form potentials in (\ref{twoForm_10D_gen}) are given by
\begin{equation}
\label{B2-C2_solu_N2}
\mathfrak{b}_{1} + i\,\mathfrak{b_2} = \frac{\cos\theta}{\sqrt{2}} e^{i\frac{\pi}{4}-i\phi}\left[ (  d\theta - \tfrac{i}{2} \sin(2\theta)\, d\phi  ) \wedge \sigma_3 - 4 \, \Delta^4 \, \sin(2\theta) \,\sigma_1 \wedge \sigma_2 \right] \ ,
\end{equation}
and the axion-dilaton matrix reads
\begin{equation}
\mathfrak{m}_{\alpha\beta} =\tfrac{1}{2} \, \Delta^2 \, \begin{pmatrix} 5-\cos(2\theta) + 2 \sin^2\theta \sin(2\phi) & 2 \sin^2\theta \cos(2\phi) \\ 2 \sin^2\theta \cos(2\phi)&   5 - \cos(2\theta) - 2 \sin^2\theta \sin(2\phi) \end{pmatrix} \ .
\end{equation}
The self-dual five-form flux takes the form
\begin{equation}
\label{F5_solu_N2}
\widetilde{F}_5 = 
4 \, \Delta^{4} \, \cos^3\theta \, (1+\star) \left[ - 3  \, \text{vol}_5 
+ \sin\theta \, d\eta \wedge \textrm{Re}\left[ \, e^{-2 i \phi} (  d\theta - \tfrac{i}{2} \sin(2\theta)\, d\phi  ) \right] \wedge \text{vol}_{3}\right] \ ,
\end{equation}
where $\,\textrm{vol}_5=\textrm{vol}_{2} \wedge \textrm{vol}_{3}$ with $\,\textrm{vol}_{2}=\sin\theta \, d\theta \wedge d\phi\,$ and $\,\textrm{vol}_{3}=-\sigma_{1} \wedge \sigma_{2} \wedge\sigma_{3}$. The radius of the external $\,\textrm{AdS}_{4}$ space-time turns out to be $\,L^2 = 1\,$ thus coinciding with the one of the $\,\mathcal{N}=4\,$ S-fold. In fact, using an effective four-dimensional supergravity description, the $\,\mathcal{N}=2\,$ and $\,\mathcal{N}=4\,$ S-folds were shown to be connected by a continuous parameter, denoted $\,\varphi\,$ in \cite{Bobev:2021yya}, leaving the AdS$_{4}$ radius $\,L\,$ unaffected and being compatible with $\,\mathcal{N}=2\,$ supersymmetry. In this manner, the $\,\varphi\,$ parameter was identified with a marginal deformation specifying a direction in a conformal manifold of $\,\mathcal{N}=2\,$ S-fold CFT's \cite{Arav:2021gra,Bobev:2021yya}. The apparent non-compactness of the parameter $\,\varphi$, at least in the four-dimensional effective description, has been the subject of recent studies and puzzles \cite{Bobev:2021yya,Cesaro:2021tna}. Working out its uplift to ten dimensions could shed some new light on this question.

\subsection{$\mathcal{N}=1$ and $\,\textrm{SU}(3)\,$ symmetric S-fold}

The $\,\mathcal{N}=1\,$ and $\,\textrm{SU}(3)\,$ symmetric S-fold with internal geometry $\,\mathcal{M}=\textrm{S}^{5}\,$ was presented in \cite{Guarino:2019oct}. Its generalisation to Sasaki--Einstein internal manifolds was discussed in \cite{Bobev:2019jbi} (see also \cite{Lust:2009mb} for a local characterisation of this type of solutions). We will take advantage of the $\,\textrm{SU}(2)$-structure of $\,\textrm{S}^5\,$ when viewed as a Sasaki--Einstein manifold, and discuss $\,\textrm{S}^5\,$ as a circle $\,\textrm{S}^{1}\,$ fibered over $\,\mathbb{CP}^2$. The coordinate on the $\,\textrm{S}^{1}\,$ is denoted $\,\phi \in [0,\,2\pi]\,$ and those on $\,\mathbb{CP}^2\,$ are
\begin{equation}
\theta \in [0,\pi] 
\hspace{5mm} , \hspace{5mm}
\alpha \in[0, 2\pi]
\hspace{5mm} , \hspace{5mm}
\beta \in[0,\,\pi]
\hspace{5mm} , \hspace{5mm}
\gamma\in[0, 4\pi] \ .
\end{equation}
Introducing a basis of one-forms on $\,\textrm{S}^{5}\,$ of the form
\begin{equation}
\label{tau_eta_definitions}
\tau_0 = d\theta
\hspace{2mm} , \hspace{2mm}
\tau_1 = \sin\theta \, \sigma_1
\hspace{2mm} , \hspace{2mm}
\tau_2 = \sin\theta \, \sigma_2
\hspace{2mm} , \hspace{2mm}
\tau_3 = \tfrac{1}{2}  \sin(2\theta) \, \sigma_3
\hspace{3mm} , \hspace{3mm}
\boldsymbol{\eta} = d\phi \,+\, \sin^2\theta \, \sigma_3 \ ,
\end{equation}
with $\,\sigma_{1,2,3}\,$ given in (\ref{SU2-inv-forms}) and $\,\boldsymbol{\eta}\,$ being the real one-form of the $\,\textrm{SU}(2)$-structure, the internal part of the metric in (\ref{metric_10D_gen}) reads
\begin{equation}
\label{metric_10D_SU3}
ds_{6}^2 = \frac{5\sqrt{5}}{18} \left( \, \frac{2}{3} \, d\eta^2 + ds^2_{\mathbb{CP}^{2}} + \frac{6}{5} \, \boldsymbol{\eta}^2 \, \right) 
\hspace{8mm} \textrm{ with } \hspace{8mm}
ds^2_{\mathbb{CP}^{2}}=\displaystyle\sum_{a=0}^{3}{\tau_a}^2 \ ,
\end{equation}
and the warping factor takes the constant value
\begin{equation}
\Delta = \frac{5}{3\sqrt{6}} \ .
\end{equation}
The $\eta$-independent two-form potentials and axion-dilaton matrix in (\ref{twoForm_10D_gen}) are given by
\begin{equation}
\label{B2-C2_SU3}
\mathfrak{b}_{1} - i \,\mathfrak{b}_2  = \frac{1}{\sqrt{6}} \, e^{i \frac{\pi}{4}}  \, \boldsymbol{\Omega}
\hspace{10mm} \textrm{ and } \hspace{10mm}
\mathfrak{m}_{\alpha\beta} = \mathbb{I}_{\alpha\beta}\ ,
\end{equation}
with $\,\boldsymbol{\Omega}=e^{3 i \phi} \, (\tau_0 + i\,\tau_3)\wedge(\tau_1+i\,\tau_2)\,$ being the complex $\,(2,0)$-form of the $\,\textrm{SU}(2)$-structure. The two-form $\,\boldsymbol{\Omega}\,$ is charged under the $\,\textrm{U}(1)_{\phi}\,$ isometry of the fibration $\,\textrm{S}^5 \sim \mathbb{CP}^2 \rtimes \textrm{S}^{1}$. Therefore, the complex two-form potential in (\ref{B2-C2_SU3}) breaks this $\,\textrm{U}(1)_{\phi}\,$ isometry of the metric (\ref{metric_10D_SU3}). The self-dual five-form reads
\begin{equation}
\label{F5tilde_SU3}
\widetilde{F}_5 = - 3 \, (1+\star)  \, \text{vol}_5 \ , 
\end{equation}
with $\,\text{vol}_5=\textrm{vol}_{\mathbb{CP}^{2}} \wedge \boldsymbol{\eta}\,$ and $\,\textrm{vol}_{\mathbb{CP}^{2}}=-\tau_{0} \wedge \tau_{1} \wedge \tau_{2} \wedge \tau_{3}$. Lastly, the $\,\textrm{AdS}_4\,$ radius takes the value $\,L^2 = 5^{\frac{5}{2}} \cdot  3^{-3} \cdot 2^{-1}$.

\subsection{$\mathcal{N}=0$ and $\,\textrm{SO}(6)\,$ symmetric S-fold}

This S-fold first discussed in \cite{Guarino:2019oct} is the simplest one. The internal $\,\textrm{S}^{5}\,$ is round and features its largest possible $\,\textrm{SO}(6)\,$ symmetry, although no supersymmetry is preserved by the solution. The internal metric, constant warping factor and self-dual five-form flux are given by
\begin{equation}
\label{10D_soluN0}
ds_{6}^2 =  \tfrac{1}{2\sqrt{2}} \,  d\eta^2 +  \tfrac{1}{\sqrt{2}} \, {ds}^2_{\textrm{S}^5}
\hspace{5mm},\hspace{5mm} 
\Delta = \frac{1}{\sqrt{2}} \hspace{5mm},\hspace{5mm}
\widetilde{F}_5 = \mp \,  4 \, (1+\star) \, \text{vol}_5 \ ,
\end{equation}
with $\,\text{vol}_5\,$ being the volume form of the round $\,\textrm{S}^{5}$ of unit radius. The $\,\eta$-independent two-forms and axion-dilaton matrix in (\ref{twoForm_10D_gen}) read
\begin{equation}
\label{B2-C2_alt}
\mathfrak{b}_{i} = 0
\hspace{10mm} \textrm{ and } \hspace{10mm}
\mathfrak{m}_{\alpha\beta} = \mathbb{I}_{\alpha\beta} \ .
\end{equation}
Finally, the $\textrm{AdS}_4$ radius takes the value $\,L^2=3 \cdot  2^{-\frac{3}{2}}$. The analysis of scalar fluctuations performed in \cite{Guarino:2019oct,Guarino:2020gfe} at the four-dimensional supergravity level showed that this non-supersymmetric S-fold features perturbatively instabilities: various scalar modes have a tachyonic mass $\,m^2L^2=-3\,$ lying below the Breitenlohner--Freedman (BF) bound \cite{Breitenlohner:1982bm} for stability in AdS$_{4}$.

\section{Flat deformations of type IIB S-folds: four-dimensional picture}

In the previous section we have presented four classes of S-fold solutions with $\,\mathcal{N}=0 \,\&\,\textrm{SO}(6)\,$, $\,\mathcal{N}=1 \,\&\,\textrm{SU}(3)\,$, $\,\mathcal{N}=2 \,\&\,\textrm{SU}(2) \times \textrm{U}(1)\,$ and $\,\mathcal{N}=4 \,\&\,\textrm{SO}(4)\,$ (super) symmetries. The symmetries of these solutions nicely match with the largest symmetric SYM interfaces classified in \cite{DHoker:2006qeo}. This is not so surprising as SYM interfaces are dual to Janus solutions and Janus solutions can be connected to S-folds upon the limiting procedure presented in \cite{Bobev:2020fon}. However, it was stated in \cite{DHoker:2006qeo} that other interfaces with the same amount of supersymmetry but less internal symmetries may be viewed as perturbations of the former by BPS operators that further break the internal symmetry. This immediately poses the question of what are such marginal deformations from the gravity side and, moreover, whether they survive the S-fold limit. 

In order to search for such deformations of the S-fold backgrounds we will move to their four-dimensional realisation as AdS$_{4}$ vacua of the $\,[\textrm{SO}(1,1) \times \textrm{SO}(6)] \ltimes \mathbb{R}^{12}\,$ maximal supergravity. Within this context, searching for marginal deformations in the conjectured S-fold CFT's amounts to search for flat directions in the scalar potential of the gauged supergravity that leave the AdS$_{4}$ radius $\,L\,$ unaffected. We will generically refer to these as \textit{flat deformations}. Such flat deformations must correspond to specific directions in the scalar manifold of the gauged supergravity. Being a maximal supergravity in four dimensions, the scalar geometry is fixed and identified with the maximally non-compact coset space $\,\textrm{E}_{7(7)}/\textrm{SU}(8)$. Therefore, any point $\,\phi\,$ in the scalar field space is identified with an element in the coset space, namely, $\,\mathcal{V}(\phi) \in \textrm{E}_{7(7)}/\textrm{SU}(8)$. In particular, each of the AdS$_{4}$ vacua representing the S-fold solutions in section~\ref{sec:S-fold_solutions} corresponds to some specific point $\,\phi_{0}\,$ in field space and, therefore, has an associated coset element $\,\mathcal{V}(\phi_{0}) \in \textrm{E}_{7(7)}/\textrm{SU}(8)\,$.

We can now take advantage of the $\,\textrm{E}_{7(7)}/\textrm{SU}(8)\,$ coset structure of the scalar manifold in the maximal $\,\textrm{D}=4\,$ supergravities. Coset spaces are homogeneous spaces so any two points in the scalar geometry are connected by a non-compact $\,\textrm{E}_{7(7)}\,$ transformation. Since we are searching for flat deformations of the S-folds discussed above, we have to search for flat directions in the scalar potential parameterised by some scalars that we collectively denote $\,\chi\,$. Then the coset structure of the scalar space allows us to parameterise a general flat deformation $\,\chi\,$ in terms of a minimal replacement
\begin{equation}
\label{Vprime}
\mathcal{V}_{0} \rightarrow \mathcal{V}_\chi \, \mathcal{V}_0 \ ,
\end{equation}
where $\,\mathcal{V}_0\,$ denotes the coset element at the \textit{undeformed} S-fold AdS$_{4}$ vacuum and $\,\mathcal{V}_\chi \,$ denotes the non-compact $\,\textrm{E}_{7(7)}\,$ transformation induced by the constant flat deformation $\,\chi\,$. We see from (\ref{Vprime}) that the undeformed S-fold $\,\mathcal{V}_0 \,$ can be thought of as a seed solution from which we can generate a larger class of solutions upon acting with $\,\mathcal{V}_\chi\,$.

However, this is not the end of the story. The Lagrangian of a given gauged supergravity is fully encoded into an object $\,\Theta\,$ called the \textit{embedding tensor} which transforms in the $\,\bf{912}\,$ irrep of $\,\textrm{E}_{7(7)}\,$  \cite{deWit:2007mt}. We will denote the embedding tensor of our theory $\,\Theta^{[\textrm{SO}(1,1) \times \textrm{SO}(6)] \ltimes \mathbb{R}^{12}}$. Then, due to the formally $\,\textrm{E}_{7(7)}\,$ covariance of the maximal $\,\textrm{D}=4\,$ supergravities \cite{deWit:2007mt}, we can alternatively view the flat-deformed vacuum at $\,\mathcal{V}_\chi  \mathcal{V}_0\,$ in the theory $\,\Theta^{[\textrm{SO}(1,1) \times \textrm{SO}(6)] \ltimes \mathbb{R}^{12}}\,$ as an undeformed vacuum at $\,\mathcal{V}_0\,$ in a \textit{different} theory specified by a new embedding tensor $\,\widetilde{\Theta}\,$ of the form    
\begin{equation}
\label{tildeTheta}
\widetilde{\Theta} = \mathcal{V}_{\chi} \star  \Theta^{[\textrm{SO}(1,1) \times \textrm{SO}(6)] \ltimes \mathbb{R}^{12}} =  \Theta^{[\textrm{SO}(1,1) \times \textrm{SO}(6)] \ltimes \mathbb{R}^{12}} + \delta\Theta \ ,
\end{equation}
where $\,\star\,$ denotes the action of the $\,\textrm{E}_{7(7)}\,$ element $\, \mathcal{V}_{\chi}\,$ in (\ref{Vprime}) on the embedding tensor $\, \Theta \in \textbf{912}\,$ of the original theory, in our case, the $\,[\textrm{SO}(1,1) \times \textrm{SO}(6)] \ltimes \mathbb{R}^{12}\,$ maximal supergravity. The new theory encoded in the new embedding tensor $\,\widetilde{\Theta}\,$ will generically differ from the original one encoded in the original embedding tensor $\, \Theta^{[\textrm{SO}(1,1) \times \textrm{SO}(6)] \ltimes \mathbb{R}^{12}}$. Adopting this viewpoint will facilitate the study of flat deformations: studying them at the level of the embedding tensor is much simpler than doing it at the level of the scalar sector.

From the discussion above, one then arrives at the following necessary conditions for a flat deformation to exist
\begin{equation}
\label{V=V=V}
V(\widetilde{\Theta} \, , \, \mathcal{V}_0) = V(\Theta \,,\, \mathcal{V}_\chi  \mathcal{V}_0) = V(\Theta \, ,\,\mathcal{V}_0)\ .
\end{equation}
The first equality is just a consequence of the formally $\textrm{E}_{7(7)}$-covariance of the maximal $\,\textrm{D}=4\,$ supergravities previously discussed. The second equality implies that the deformation is flat, namely, that the value of the scalar potential at the deformed and undeformed vacua in the original theory is the same. An explicit computation of $\,V(\widetilde{\Theta} \, , \, \mathcal{V}_0)\,$ in (\ref{V=V=V}) yields the scalar potential of the original $\,\Theta^{[\textrm{SO}(1,1) \times \textrm{SO}(6)] \ltimes \mathbb{R}^{12}}\,$ theory together with two additional contributions. The first one is the scalar potential $\,V(\delta \Theta,\,\mathcal{V}_0)$ of the would-be $\,\delta\Theta\,$ theory. The second one corresponds with a mixed term involving both $\,\Theta^{[\textrm{SO}(1,1) \times \textrm{SO}(6)] \ltimes \mathbb{R}^{12}}\,$ and $\,\delta \Theta\,$. It was then shown in \cite{Guarino:2021hrc} that a class of $\,\delta\Theta$-theories for which these two contributions vanish identically is given by the Cremmer--Scherk--Schwarz (CSS) \textrm{flat gaugings} of \cite{Cremmer:1979uq}\footnote{For this class of CSS gauging, the necessary conditions in (\ref{V=V=V}) to have a flat deformation become also sufficient.}. In \textit{ungauged} four-dimensional supergravity, a flat gauging $\,\delta\Theta^{\textrm{CSS}}\,$ is parameterised by a generic element $\,\chi\,$ of the maximally compact subalgebra $\,\mathfrak{usp}(8) \subset \mathfrak{e}_{6(6)}\,$ of the five-dimensional duality group $\,\textrm{E}_{6(6)}$. In our case, due to the non-abelian $\,\textrm{SO}(6)\,$ factor in the $\,[\textrm{SO}(1,1) \times \textrm{SO}(6)] \ltimes \mathbb{R}^{12}\,$ \textit{gauged} four-dimensional supergravity, only the Cartan subalgebra of $\,\mathfrak{so}(6) \subset \mathfrak{usp}(8)\,$ can be consistently implemented as a flat deformation. More concretely, there always exists a $\,\mathcal{V}_\chi\,$ such that
\begin{equation}
\widetilde{\Theta} = \mathcal{V}_{\chi} \star  \Theta^{[\textrm{SO}(1,1) \times \textrm{SO}(6)] \ltimes \mathbb{R}^{12}} = \Theta^{[\textrm{SO}(1,1) \times \textrm{SO}(6)] \ltimes \mathbb{R}^{12}} + \delta\Theta^{\textrm{CSS}}(\chi) \ ,
\end{equation}
with $\,\chi \equiv \chi_{i}{}^{j}\,$ being valued in the Cartan subalgebra of $\,\mathfrak{so}(6)$. In summary, turning on a specific element $\,\chi\,$ in the Cartan subalgebra of the residual symmetry group $\,\textrm{G}_{0}\,$ and performing the minimal replacement in (\ref{Vprime}) still extremises the scalar potential of the original theory and therefore can be seen as a solution generating technique. However, this will break the $\,\textrm{G}_{0}\,$ symmetry down to the subgroup commuting with $\,\chi\,$. Therefore, a generic choice of $\,\chi\,$ will fully break the $\,\textrm{G}_{0}\,$ symmetry of the undeformed S-fold down to its maximal Cartan subgroup $\,\textrm{U}(1)^{\textrm{Rank}(\textrm{G}_{0})}$. This construction is schematically summarised in the bottom part of Figure~\ref{fig:Schematic_flat_deformations}.

Notice that the above arguments do not rely on $\,\textrm{G}_{0}\,$ being identified with an R-symmetry or a flavour symmetry group in the dual S-fold CFT (we will attach labels R and F to emphasise this aspect). As a consequence, one can break either type of symmetries with the flat deformation mechanism we just described. For example, when considering the $\,\mathcal{N}=2\,$ and $\,\textrm{SU}(2)_{\textrm{F}} \times \textrm{U}(1)_{\textrm{R}}\,$ S-fold in section~\ref{sec:N=2_solution}, there are two distinct flat deformations: one breaks the $\,\textrm{SU}(2)_{\textrm{F}}\,$ flavour group down to its $\,\textrm{U}(1)_{\textrm{F}}\,$ Cartan and the other breaks supersymmetry as it is associated with the $\,\textrm{U}(1)_{\textrm{R}}\,$ R-symmetry factor. In the case of the $\,\mathcal{N}=4\,$ and $\,\textrm{SO}(4)_{\textrm{R}}\,$ symmetric S-fold of section~\ref{sec:N=4_solution}, the entire symmetry group maps into the R-symmetry of the dual $\,\mathcal{N}=4\,$ S-fold CFT. We will discuss this case in more detail below. Lastly, it is worth emphasising that flat deformations provide us with a controlled mechanism of supersymmetry breaking in the context of type IIB S-folds.

\subsection{Deforming the $\,\mathcal{N}=4\,$ and $\,\textrm{SO}(4)\,$ symmetric S-fold}

\begin{figure}[t]
\centering
\includegraphics[scale=0.6]{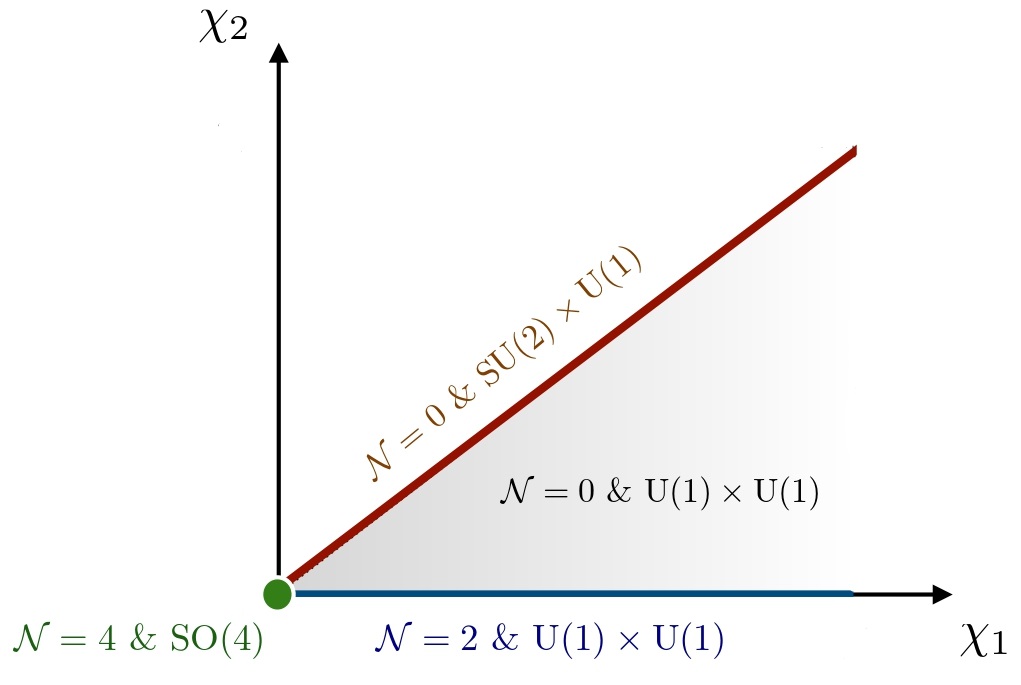}
\caption{Patterns of (super) symmetry breaking induced by the flat deformations $\,(\chi_{1},\chi_{2})\,$ for the $\,\mathcal{N}=4\,$ and $\,\textrm{SO}(4)\,$ symmetric S-fold.}
\label{fig:N4_modulispace_4D}
\end{figure}

Let us consider the $\,\mathcal{N}=4\,$ and $\,\textrm{G}_{0}=\textrm{SO}(4)\,$ symmetric S-fold solution of section~\ref{sec:N=4_solution}. A generic flat deformation $\,\chi\,$ in the Cartan subalgebra of $\, \mathfrak{so}(4)\,$ will break the S-fold symmetry from $\,\textrm{SO}(4)\,$ to its maximal Cartan subgroup $\,\textrm{U}(1)^2$. Let us choose, in the appropriate basis, the constant flat deformation to be
\begin{equation} 
\label{chi_mat}
\chi_{i}{}^{j} = \sqrt{6} \,  \left(
\begin{array}{cccccc}
0 & 0 & 0 & 0 & 0 & 0 \\
0 & 0 & 0 & 0 & 0 & 0 \\
0 & 0 & 0 & 0 & \chi_{1} & \chi_{2} \\
0 & 0 & 0 & 0 & \chi_{2} & \chi_{1} \\
0 & 0 & -\chi_{1}  & -\chi_{2} & 0 & 0 \\
0 & 0 & -\chi_{2} & -\chi_{1}  & 0 & 0
\end{array}
\right) \in \mathfrak{so}(2)^2  \subset \mathfrak{so}(4) \ ,
\end{equation}
so that two flat deformations $\,\chi_{i}=(\chi_{1},\chi_{2})\,$ are allowed. Modding out by the discrete symmetries of the system, namely, $\,\chi_{i} \leftrightarrow -\chi_{i}\,$ and $\,\chi_{1} \leftrightarrow \chi_{2}\,$, one finds the patterns of (super) symmetry breaking displayed in Figure~\ref{fig:N4_modulispace_4D}.

An analysis of the scalar fluctuations about this AdS$_{4}$ S-fold vacuum at generic values of the flat deformations $\,\chi_{i}\,$ shows that there is no scalar mode violating the BF bound. The normalised scalar spectrum was computed in \cite{Guarino:2021hrc} and is given (including Goldstone modes) by
\begin{equation}
\label{scalars_N4}
\begin{array}{lll}
m^2 L^2 &=& 10 \,\, (\times 1) \,\,\, , \,\,\, 4 \,\, (\times 2) \,\,\, , \,\,\, -2 \,\, (\times 3) \,\,\, , \,\,\, 0 \,\, (\times 32) \,\,\, , \,\,\, \chi_{i}^2 \,\, (\times 2) \ , \\[2mm]
& & (\chi_{1} \pm \chi_{2})^2 \,\, (\times 2)  \,\,\, , \,\,\,   \frac{1}{4} \, (\chi_{1} \pm \chi_{2})^2 \,\, (\times 4) \,\,\, , \,\,\, 
1+ \chi_{i}^2 \pm \sqrt{9 + 4 \,  \chi_{i}^2} \,\, (\times 2) \ , \\[3mm]
& & 1+ \frac{1}{4} \, (\chi_{1}+\chi_{2})^2 \pm \sqrt{9 + (\chi_{1}+\chi_{2})^2} \,\, (\times 2)  \ , \\[3mm]
& & 1+ \frac{1}{4} \, (\chi_{1}-\chi_{2})^2 \pm \sqrt{9 + (\chi_{1}-\chi_{2})^2} \,\, (\times 2)  \ ,
\end{array}
\end{equation}
in terms of the AdS$_{4}$ radius $\,L\,$. Therefore, this two-parameter family of AdS$_{4}$ S-fold vacua are perturbatively stable at generic values of $\,\chi_{i}\,$. It is also instructive to present the $\chi$-dependence of the gravitino masses
\begin{equation}
\label{gravitini_N4}
\begin{array}{lll}
m^2 L^2 &=& \frac{5}{2} +\frac{1}{4} \chi_{i}^2  \pm  \frac{1}{2} \sqrt{9+\chi_{i}^{2}}    \,\,\,\, (\times 2)  
\hspace{10mm} \textrm{with} \hspace{10mm}
i=1,2  \ .
\end{array}
\end{equation}
From (\ref{gravitini_N4}) it is straightforward to see that the gravitinos become massive ($m^2 L^2 >1$) when turning on the flat-deformations $\,\chi_i\,$ and that, generically, this breaks all the supersymmetries. Still some supersymmetry is preserved when one of the flat deformations vanishes (see Figure~\ref{fig:N4_modulispace_4D}).

Importantly, four-dimensional perturbative stability by no means implies higher-dimensional perturbative stability. The reason is that higher Kaluza--Klein (KK) modes could become unstable away from the supersymmetric limit $\,\chi_{i}=0\,$. This would involve a realisation of the so-called "space-invaders" scenario\footnote{This is a scenario where higher KK-modes become lighter than lower KK-modes and may produce perturbative instabilities of the higher-dimensional vacuum.}. However this is not the case for this two-parameter family of vacua as we will discuss in section~\ref{sec:remarks}. Finally, it is also worth mentioning that $\,\chi_{i}\,$ are non-compact deformation parameters from a four-dimensional perspective. Surprisingly, they turn out to be compact when understood in a higher-dimensional picture.

\section{Flat deformations of type IIB S-folds: higher-dimensional picture}

So far our discussion on flat deformations of type IIB S-folds has been restricted to their four-dimensional realisation as $\textrm{AdS}_{4}$ vacua in the maximal $\,[\textrm{SO}(1,1) \times \textrm{SO}(6)] \ltimes \mathbb{R}^{12}\,$ gauged supergravity. We will now comment on their physical interpretation in five- and ten-dimensional supergravity.

\subsection{Five-dimensional picture}
\label{sec:5D}

When oxidising the $\textrm{AdS}_{4}$ vacua describing S-folds to five dimensions, one runs into an additional complication: in five dimensions, the $\,\textrm{S}^{1}\,$ along which the type IIB fields undergo the S-duality monodromy $\,J_{k}\,$ is now part of the external five-dimensional space. This implies that the profiles for the various fields in the five-dimensional SO(6)-gauged maximal supergravity must acquire a dependence on the coordinate $\,\eta\,$ along the $\,\textrm{S}^{1}\,$. Nicely, the five-dimensional SO(6)-gauged supergravity still features the global $\,\textrm{SL}(2)\,$ S-duality of type IIB string theory. The reason for this is that the $\,\textrm{SL}(2)\,$ commutes with the gauging of the $\,\textrm{SO}(6)$ isometries of $\,\textrm{S}^{5}\,$ within the $\,\textrm{E}_{6(6)}\,$ duality group of the maximal $\,\textrm{D}=5\,$ supergravities \cite{deWit:2004nw}. This opens up the possibility to perform an S-duality twist combined with a geometric twist on the $\,\textrm{S}^{5}$. This mechanism provides a five-dimensional interpretation of the flat deformations $\,\chi_{i}{}^{j}\,$ as the parameters specifying the geometric twist along the $\,\textrm{S}^{5}$. This is the twist responsible for the geometric monodromy $\,h\,$ discussed in the introduction which causes the breaking of (super) symmetries in the S-fold backgrounds (see section~\ref{sec:10D}).

Twisted reductions of five-dimensional supergravity theories down to four dimensions have been extensively investigated starting from the seminal paper \cite{Scherk:1979zr}. They take the general form \cite{Dabholkar:2002sy}
\begin{equation}
\label{M-twist}
\phi(x^\mu,\eta) = e^{M \,\eta} \star \phi(x^\mu) \ ,
\end{equation}
where $\,\eta\,$ is the coordinate along the S$^{1}$,  $\,M \in \mathfrak{e}_{6(6)}\,$ is an element of the \textit{global} duality algebra, and $\,\phi\,$ is a generic field in the five-dimensional supergravity theory. Duality twists were originally studied within the context of five-dimensional \textit{ungauged} supergravity. Within this context, and choosing $\,M\,$ in the \emph{global} duality group $\,\textrm{E}_{6(6)}\,$ of the five-dimensional supergravity theory, the entire dependence on the $\,\eta\,$ coordinate factorises out in the reduction from five to four dimensions. Importantly, further selecting $\,M\,$ in the maximal compact subalgebra $\,\mathfrak{usp}(8) \subset \mathfrak{e}_{6(6)}\,$ yields a scalar potential in the reduced four-dimensional theory admitting Minkowski vacua \cite{Cremmer:1979uq,Andrianopoli:2002mf}. However, our five-dimensional theory is the SO(6)-gauged supergravity and, as a consequence of the gauging procedure, the global duality group $\,\textrm{E}_{6(6)}\,$ is broken to a \emph{local} $\,\textrm{SO}(6)\,$ and the \emph{global} $\,\textrm{SL}(2)\,$ S-duality group commuting with it within $\,\textrm{E}_{6(6)}$. Since the five-dimensional embedding tensor is a singlet under $\,\textrm{SO(6)}\,$, identifying the flat deformations with elements $\,M = \chi_i{}^j \subset \mathfrak{so}(6)\,$ in (\ref{M-twist}) leaves the gauging invariant. As a result, the flat deformations $\,\chi_{i}{}^{j}\,$ are expected to describe trivial twists leaving the putative five-dimensional backgrounds locally invariant.

It is also interesting to keep track of the flat deformations $\,\chi_i{}^j\,$ generically belonging to the Cartan subalgebra of $\,\mathfrak{so}(6)\,$ when oxidising the four-dimensional theory to five dimensions. As argued in \cite{Guarino:2021kyp,Guarino:2021hrc} (see also \cite{Bobev:2021rtg,Berman:2021ynm}), the deformation parameters $\,\chi_i{}^j\,$ are identified (via a Kaluza--Klein reduction) with the $\,(A_{i}{}^{j})_{\eta}\,$ components along the $\,\textrm{S}^{1}\,$ of the five-dimensional vectors spanning the $\,\textrm{SO}(6)\,$ factor in the gauge group. Following the classification of bosonic deformations of $\,\mathcal{N}=4\,$ super-Yang--Mills in terms of the auxiliary fields of the off-shell $\,\mathcal{N}=4\,$ conformal supergravity \cite{Maxfield:2016lok}, the flat deformations $\,\chi_i{}^j\,$ define non-trivial one-form deformations (Wilson loops) for the five-dimensional vector fields along the $\,\textrm{S}^{1}$. Such one-form deformations are usually discarded, \textit{i.e.}, when investigating Janus solutions, by invoking a gauge-fixing argument without much regard for large gauge transformations. In this respect, it would be interesting to oxidise the AdS$_{4}$ S-fold vacua with non-vanishing $\,\chi_i{}^j\,$ deformations to five dimensions.

\subsection{Ten-dimensional picture}
\label{sec:10D}

It was previously stated that the $\,\chi_{i}\,$ deformations turn out to be
compact parameters in a higher-dimensional picture. To illustrate this in an example, we present here the ten-dimensional uplift of the $\,(\chi_{1},\chi_{2})$-family of deformations of the $\,\mathcal{N}=4\,$ S-fold \cite{Giambrone:2021wsm}. Choosing an appropriate Cartan subalgebra of $\,\mathfrak{so}(4)\,$ to parameterise the constant deformations $\,(\chi_{1},\chi_{2})$, one finds the same ten-dimensional S-fold solution as in section~\ref{subsection:N4sol} with one crucial difference: the one-forms $\,d\varphi_i\,$ along the azimutal angles of the two-spheres $\,\textrm{S}^2_{i}\,$ in (\ref{metric_10D_solu_N4})-(\ref{spheres_S2}) get replaced by new one-forms $\,d\varphi_i + \chi_i \,d\eta$. This change of one-form basis can be \textit{locally} reabsorbed in a change of coordinates 
\begin{equation}
\varphi_i' = \varphi_i +\, \chi_i \, \eta \ .
\end{equation}
Since all the type IIB fields in the S-fold solution of section~\ref{subsection:N4sol} are independent of $\,\varphi_i$, the new backgrounds with $\,\chi_i \neq 0\,$ obtained by the minimal replacement $\,d\varphi_{i} \rightarrow d\varphi'_{i}\,$ automatically solve the equations of motion. At this stage, the reader might be tempted to conclude that the solutions with $\,\chi_i = 0\,$ and $\,\chi_i \neq 0\,$ are simply the same solution but in different coordinate patches. However, due to the periodicities $\,\varphi_i \sim \varphi_i+2\pi\,$ and $\,\eta \sim \eta + T$, there is no \textit{global} diffeomorphism connecting the two solutions unless $\,\chi_{i} = n_{i} \frac{2\pi}{T}\,$ with $\,n_{i} =  \mathbb{Z}$. This can be understood also from the fact that $\,\varphi_i'\,$ is not a globally well-defined coordinate on $\,\textrm{S}^5 \times \textrm{S}^1$. Indeed, the variable $\,\varphi'_{i}\,$ gets shifted by the quantity $\,\chi_i \, T\,$ when making a loop around $\,\textrm{S}^1$. And this quantity becomes a multiple of $\,2\pi$ only when $\,\chi_{i} = n_{i} \frac{2\pi}{T}\,$ with $\,n_{i} = \mathbb{Z}$. Only in this case, $\,\varphi_i'\,$ is a globally well-defined new azimutal angle.

There is a more general and systematic way of introducing flat deformations $\,\chi\,$ in a solution of any diffeomorphism-invariant theory involving a factorised geometry of the form $\,\mathcal{M}\,\times\, \textrm{S}^1\,$ \cite{Guarino:2021kyp}. Let us denote such a seed solution $\,\Phi_0\,$ which encodes the metric and the various $p$-form fluxes solving the equations of motion of the theory. Let us also assume that $\,\Phi_0\,$ is invariant under the action of a Lie group $\,\textrm{G}_{0}\,$ of rank $\,r$. Then, we claim that it is always possible to construct an $r$-dimensional family of solutions generically breaking $\,\textrm{G}_{0}\,$ down to its maximal Cartan subgroup $\,\textrm{U}(1)^r$. To do so, we first choose an element $\,h \in \textrm{G}_{0}\,$ which, without loss of generality, can be chosen in a Cartan subgroup of $\,\textrm{G}_{0}$. From the element $\,h\,$, we build the so-called \textit{mapping torus}
\begin{equation}
\label{mapping_torus}
T(\mathcal{M})_h = \frac{\mathcal{M} \times [0,\,T]}{(p,\,0)\sim(h\cdot p,\,2\pi)} \ ,
\end{equation}
for all the points $\,p \in \mathcal{M}$. The quotient introduced in (\ref{mapping_torus}) is just a way of encoding different ``boundary conditions'' (periodicities) for the angular coordinates $\,\varphi_{i}\,$ on $\,\mathcal{M}\,$ which are associated with commuting isometries of the seed solution $\,\Phi_{0}$. This amounts to introduce a non-trivial $\,\textrm{SO}(6)\,$ monodromy $\,h\,$ when moving around the $\,\textrm{S}^1\,$ in the factorised geometry $\,\mathcal{M}\times \textrm{S}^1$.

To construct the new solutions with non-zero flat deformations, $\,\chi_{i} \neq 0\,$, we first project the seed solution $\,\Phi_0\,$ defined on $\,\mathcal{M}\times \textrm{S}^1\,$ to a solution on the mapping torus $\,T(\mathcal{M})_h\,$. This is done by the canonical projection
\begin{equation}
\pi : \mathcal{M}\times \textrm{S}^1 \xrightarrow[]{} \mathcal{M}\times [0,\,T] \xrightarrow[]{\sim} T(\mathcal{M})_h\ .
\end{equation}
If $\,h\neq \mathbb{I}\,$, this projection is \emph{not} a diffeomorphism: while it is locally well-defined, it is not globally well-defined. Still, the projection $\,\pi(\Phi_0)\,$ is well-defined as the action of $\,h\in \textrm{G}_0\,$ on $\,\Phi_0\,$ is trivial by definition. Nonetheless, $\,\pi(\Phi_0)\,$ continues being a solution of the theory as the equations of motion are local. However the symmetry group of $\,T(\mathcal{M})_h\,$ gets reduced to those elements of $\,\textrm{G}_{0}\,$ commuting with $\,h\,$. Otherwise, their action is not globally well-defined.

\begin{figure}[t]
\centering
\includegraphics[width=\textwidth]{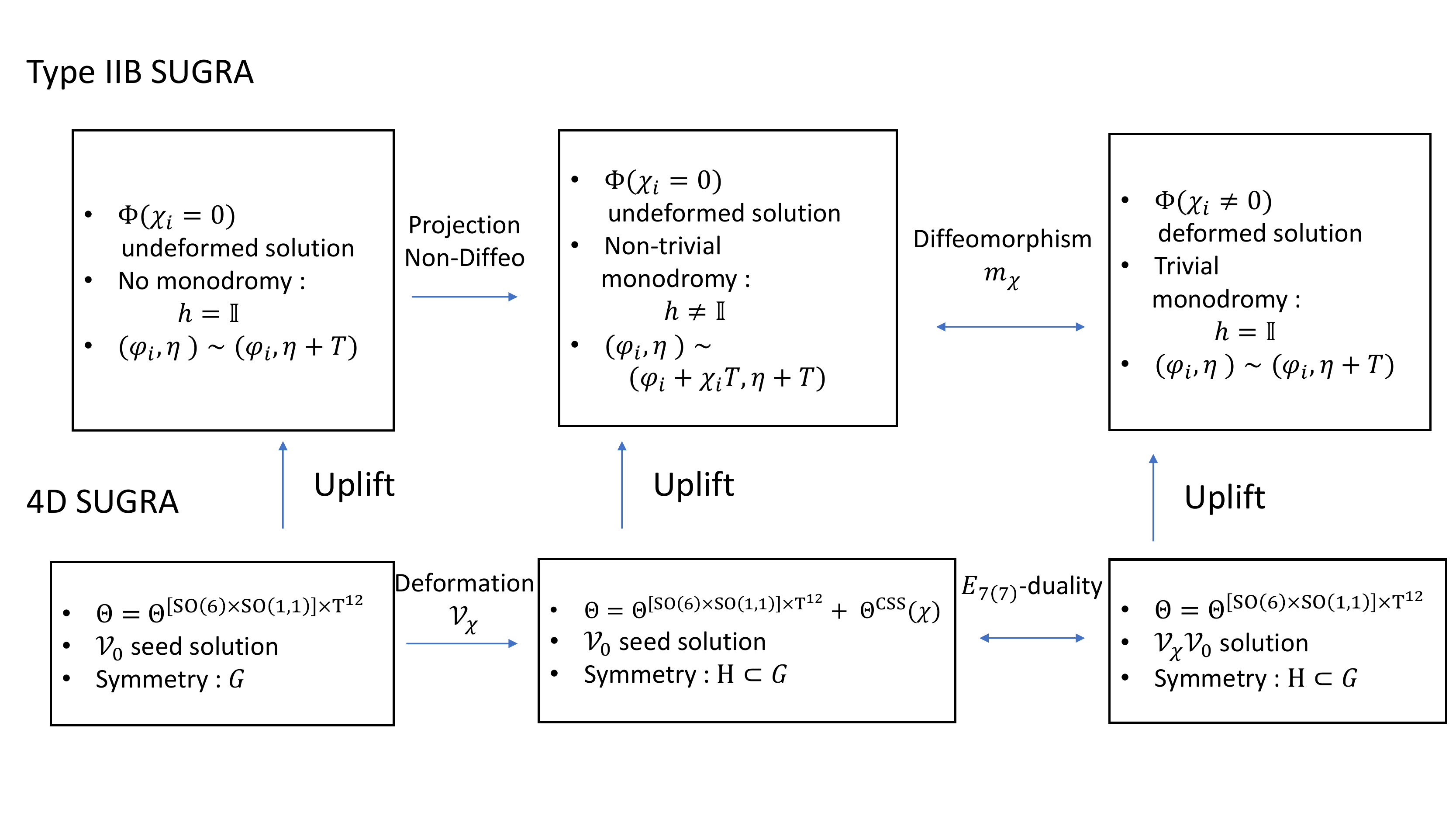}
\caption{Schematic representation of the flat deformations from four- and ten-dimensional perspectives.}
\label{fig:Schematic_flat_deformations}
\end{figure}

Let us illustrate the above construction when selecting the $\,\mathcal{N}=4\,$ and $\,\textrm{SO}(4)\,$ symmetric S-fold as the seed solution $\,\Phi_{0}$. In this case we choose $\,h= \exp(T \, \chi)$ with $\,\chi \in {\mathfrak{u}(1)}^2 \subset \mathfrak{so}(4)$. We also introduce coordinates $\,\varphi'_i\,$ on the mapping torus $\,T(\mathcal{M})_h\,$ which get shifted by $\,\chi_i \in \mathfrak{u}(1)\,$ and are subject to the identifications
\begin{equation}
\label{boundary_cond}
(\varphi'_i,\,\eta) \sim (\varphi'_i+\chi_i \, T,\,\eta+T) \hspace{5mm}\text{and}\hspace{5mm} (\varphi'_i,\,\eta) \sim (\varphi'_i+2\pi,\,\eta)\ .
\end{equation}
The projection from $\,\textrm{S}^5 \times \textrm{S}^1\,$ to the mapping torus $\,T(\textrm{S}^5)_h\,$ is trivial: it is the replacement $\,\varphi_{i} \rightarrow \varphi_{i}'$. Using the $\,\varphi_{i}'\,$ coordinates, the type IIB fields are $\,\chi_i$-independent and the flat deformations are totally encoded in the choice of periodicities in (\ref{boundary_cond}). However, there is an alternative coordinate system in which the deformations are encoded in the type IIB fields and not in the monodromy $\,h$. Let us consider a diffeomorphism of the form
\begin{equation}
\label{diffeomorphism_example}
m_\chi : T(\mathcal{M})_h \rightarrow \mathcal{M}\times \textrm{S}^1 : (p,\,\eta) \rightarrow (e^{\chi \eta} \cdot p,\,\eta) \ .
\end{equation}
It can be verified that (\ref{diffeomorphism_example}) is indeed a well-defined diffeomorphism, even globally, due to the different choice of coordinates identification when moving around the $\,\textrm{S}^1$. This corresponds, locally, to a change of coordinates of the form
\begin{equation}
(\varphi_{i}',\,\eta) \rightarrow (\varphi'_{i}+\chi_i\,\eta,\,\eta) \ , 
\end{equation}
where $\,\varphi_{i} \sim \varphi_{i} + 2\pi\,$ and $\,\eta \sim \eta + T\,$. Since $\,m_\chi\,$ in (\ref{diffeomorphism_example}) is globally a well-defined diffeomorphism, this implies that solutions depending explicitly on the $\,\chi_i\,$ deformations and involving an internal manifold with \emph{trivial} monodromy $\,h\,$ are equivalent to solutions with no explicit dependence on the $\,\chi_i\,$ deformations but involving an internal manifold with \emph{non-trivial} monodromy $\,h$. For the confused reader (we were too), this is summarised in Figure~\ref{fig:Schematic_flat_deformations}.

Although a full proof of the bijection between four-dimensional flat deformations and the ten-dimensional construction we just presented has not yet been formally established, we have verified its validity in the various S-fold solutions presented in section~\ref{sec:S-fold_solutions} (see \cite{Guarino:2021kyp}). Whenever valid, it has strong implications for the conjectured S-fold CFT's. Recall that the flat deformations $\,\chi_{i}\,$ do not change the AdS$_{4}$ radius $\,L\,$ of the S-fold solutions and therefore, at least in the large $\,N\,$ limit, they correspond to marginal deformations of the dual S-fold CFT's. It is of relevance to understand what are the global properties of the conformal manifold of S-fold CFT's. For example, whether or not such a conformal manifold is compact. A quick inspection of the scalar and gravitino normalised masses in the four-dimensional spectra (\ref{scalars_N4}) and (\ref{gravitini_N4}) is enough for us to conclude that these spectra are \textit{not} periodic under a shift $\,\chi_{i} \rightarrow \chi_{i} + \frac{2\pi}{T}$. However, the ten-dimensional geometric interpretation we provided in terms of the mapping torus suggests that the physics of S-folds should be invariant under such a shift.\footnote{Recall also that the mapping torus was built upon the group element $\,h= \exp(T \chi) \in \textrm{G}_{0}\,$ and not upon $\,\chi \in \mathfrak{g}_{0}\,$ so that $\,\chi_i \rightarrow \chi_i + \frac{2\pi}{T}\,$ was indeed a symmetry of the ten-dimensional backgrounds.} The resolution of this paradox comes from the study of Kaluza--Klein (KK) modes in the S-fold backgrounds (see \cite{Giambrone:2021zvp} for the KK spectrometry of the $\,\chi$-deformed $\,\mathcal{N}=2\,$ and $\,\textrm{U}(1)^2\,$ symmetric S-folds of \cite{Guarino:2020gfe}). It is usually assumed that modes higher up in the KK tower come along with a larger mass than the lower KK modes, and thus can be safely ignored. This intuition is however not always correct. When $\,\chi_{i}$-deforming the $\,\mathcal{N}=4\,$ and $\,\textrm{SO}(4)\,$ symmetric S-fold, a detailed mass spectrometry of the full tower of KK modes showed that, at $\,\chi_i = \frac{2\pi}{T}$, certain modes at a higher KK level get the same masses as certain modes at $\,\chi_i = 0\,$ belonging to a lower KK level \cite{Giambrone:2021wsm}. In other words, modes sitting at different KK levels get reshuffled amongst themselves but the full (higher-dimensional) KK spectrum becomes periodic under the shift $\,\chi_i \rightarrow \chi_i + \frac{2\pi}{T}\,$. Therefore, flat deformations $\,\chi_{i}\,$ manifest themselves as non-compact deformations only within the realm of four-dimensional supergravity.

\section{Flat deformations of type IIB S-folds: holographic picture}

At the present time, it is fair to say that most of the holographic aspects of the various type IIB S-folds discussed above are yet to be understood. An interesting proposal has been made in \cite{Assel:2018vtq} for the $\,\mathcal{N}=4\,$ S-fold with SO(4) symmetry originally presented in \cite{Inverso:2016eet}. According to this proposal, the dual S-fold CFT would appear as the effective IR description of a three-dimensional $\,\textrm{T}[\textrm{U}(N)]\,$ theory of the type constructed in \cite{Gaiotto:2008ak} where a diagonal subgroup of the $\,\textrm{U}(N)^2\,$ flavour group is gauged using an $\,\mathcal{N} = 4\,$ vector multiplet. The hyperbolic monodromy $\,J_k \in \textrm{SL}(2,\,\mathbb{Z})\,$ is then induced by a Chern--Simons term at level $\,k\,$ that is also turned on.

An effective $\,\mathcal{N}=4\,$ superpotential $\,W_{\textrm{eff}} = (2\pi/k) \, {\rm Tr}(\mu_H\,\mu_C)$, where $\,\mu_H\,$ and $\,\mu_C\,$ are respectively complex scalars describing the Higgs and Coulomb branches of the $\,\textrm{ T}[\textrm{U}(N)]\,$ theory, was proposed in \cite{Bobev:2021yya} to be a marginal operator in the IR. Moreover, it was further argued that a shift $\,W_{\textrm{eff}} \rightarrow W_{\textrm{eff}} \, + \, \lambda \, {\rm Tr}(\mu_H\,\mu_C)\,$ with $\,\lambda \in \mathbb{C}\,$ would represent an exactly marginal deformation breaking $\,\mathcal{N} = 4\,$ down to $\,\mathcal{N}=2\,$ \cite{Bobev:2021yya}\footnote{Only one of the real components of $\,\lambda \in \mathbb{C}\,$ corresponds to a $\chi$-like flat deformation of the type discussed in this proceeding. The remaining real component, denoted $\,\varphi\,$ in \cite{Bobev:2021yya}, does not belong to the class of $\chi$-like deformations discussed here. A higher-dimensional interpretation of $\,\varphi\,$ as well as other aspects like its compact versus non-compact nature are currently under debate/investigation (see discussion below eq.(\ref{F5_solu_N2})).\label{footnote_varphi}}. Building upon these ideas, it was subsequently suggested in \cite{Giambrone:2021wsm} that adding an additional $\chi$-like flat deformation would also correspond to an exactly marginal deformation of the three-dimensional Lagrangian of the form $\,\partial_\alpha \mathcal{O} \, \partial^\alpha \bar{\mathcal{O}}\,$ with $\,\mathcal{O}\equiv  {\rm Tr}(\mu_H\,\bar{\mu}_C)\,$ with $\,\partial_\alpha\,$ denoting the partial derivatives with respect to (real) scalar fields. The $\,\partial_\alpha \mathcal{O} \, \partial^\alpha \bar{\mathcal{O}}\,$ deformation is not a holomorphic deformation of the superpotential and therefore would break all the supersymmetries. This was used as evidence for the existence of non-supersymmetric conformal manifolds in \cite{Giambrone:2021wsm}.

Additional insights into the holographic aspects of the S-fold backgrounds discussed in this proceeding have also come from the study of holographic RG-flows on the D3-brane \cite{Guarino:2021kyp,Arav:2021tpk}. Using the effective four-dimensional $\,[\textrm{SO}(1,1) \times \textrm{SO}(6)] \ltimes \mathbb{R}^{12}\,$ gauged supergravity, all the S-fold backgrounds were shown to be connected with a deformation of the D3-brane solution, namely, a deformation of the $\,\textrm{AdS}_{5} \times \textrm{S}^{5}\,$ background, via a holographic RG-flow. In other words, all the conjectured S-fold SCFT's can be viewed as IR fixed points of a supersymmetric RG-flow that starts from a deformation of $\,\mathcal{N}=4\,$ super Yang--Mills in the UV. This UV regime was further characterised in \cite{Guarino:2021kyp} and found to correspond to an \textit{anisotropic} deformation of $\,\mathcal{N}=4\,$ SYM. This is in agreement with the five-dimensional supergravity fields developing a dependence on the $\,\eta\,$ coordinate along $\,\textrm{S}^1$, which becomes a spatial coordinate at the deformed field theory living at the boundary.

The above RG-flows were also analysed in \cite{Guarino:2021kyp} from a ten-dimensional perspective aiming at shedding some new light on their holographic interpretation. In this manner, the anisotropic deformation of $\,\textrm{SYM}\,$ was interpreted in a purely geometric manner and connected to the locally geometric $\,\textrm{SL}(2)\,$ S-duality twist matrix $\,A(\eta)\,$ in (\ref{A-twist}) generating the S-fold background. Moreover, a set of five-dimensional one-form deformations $\,(\mathcal{F}_{(1)}{}^{\alpha}\,,\,\mathcal{F}_{(1)\alpha\beta}\,,\,\widetilde{\mathcal{F}}_{(1)})\,$ were identified as the sources of anisotropy, thus coming in $\,\textrm{SL}(2)\,$ representations. A more detailed characterisation of the operators triggering such an anisotropy in an $\,\textrm{SL}(2)\,$ covariant context is yet to be worked out. Lastly, holographic RG-flows between S-fold CFT's were also explicitly constructed in \cite{Guarino:2021kyp,Arav:2021gra}.

\section{Remarks on stability and future directions}
\label{sec:remarks}

Arrived at this point in our stroll through the physics of type IIB S-folds, there is the obvious question of whether the flat deformations $\,{\chi_{i}}^{j}\,$ induce instabilities in the cases where they fully break supersymmetry. This would cause the decay of the non-supersymmetric S-fold solutions as predicted by the AdS swampland conjecture \cite{Ooguri:2016pdq}.

As anticipated in the main text, perturbative instabilities do not occur for the two-parameter family of non-supersymmetric and $\,\textrm{U}(1)^2\,$ symmetric S-folds obtained upon turning on flat deformations in the original $\,\mathcal{N}=4\,$ and $\,\textrm{SO}(4)\,$ symmetric S-fold of \cite{Inverso:2016eet}. This was explicitly shown in \cite{Giambrone:2021zvp} by performing a fully detailed Kaluza-Klein spectrometry analysis of such a two-parameter family of vacua. In \cite{Giambrone:2021zvp}, the full Kaluza-Klein spectrum of the \textit{undeformed} $\,\mathcal{N}=4\,$ S-fold with $\,\chi_{i}{}^{j}=0\,$ was shown to arrange itself into long graviton multiplets. The conformal dimension of the spin 2 highest-weight state in each supermultiplet was found to be
\begin{equation}
\label{Delta_N=4}
 \Delta = \frac32 + \frac12 \sqrt{9 + 2\ell(\ell+4) + 4 \sum_{i=1,2}\ell_i(\ell_i+1)+ 2\left(\frac{2 n \pi}{T}\right)^2} \ ,
\end{equation}
where $\,\ell\,$ is the Kaluza-Klein level on the $\,\textrm{S}^5$, $\,n\,$ is the Kaluza-Klein level on the $\,\textrm{S}^1\,$ and $\,(\ell_1,\ell_2)\,$ denote the $\,\textrm{SO}(4) \sim \textrm{SU}(2) \times \textrm{SU}(2)\,$ spin representation of the highest-weight spin 2 state. It was subsequently shown in \cite{Giambrone:2021wsm} that turning on the flat deformation $\,\chi_{i}{}^{j}\,$ in (\ref{chi_mat}) simply amounts to a shift of the form
\begin{equation}
\label{shift}
 \frac{2n\pi}{T} \longrightarrow \frac{2n\pi}{T} + \sum_{i=1,2} j_i \, \chi_i  \ ,
\end{equation}
in the conformal dimensions (\ref{Delta_N=4}). The labels $\,(j_1,j_2)\,$ in (\ref{shift}) correspond to the charges of the respective fields under the appropriate $\,\textrm{U}(1) \times \textrm{U}(1)\,$ Cartan subgroup of $\,\textrm{SO}(4)\,$. As a direct consequence of (\ref{Delta_N=4})-(\ref{shift}), the conformal dimensions (and therefore the masses) of the various fields in the Kaluza-Klein tower are bounded from below by the fields in the four-dimensional supergravity. Since there were no scalars violating the BF bound in the four-dimensional spectrum of (\ref{scalars_N4}), the non-supersymmetric and $\,\textrm{U}(1)^2\,$ symmetric S-folds are perturbatively stable also in higher-dimensions.

A more subtle issue is to assess the non-perturbative stability of the non-supersymmetric and $\,\textrm{U}(1)^2\,$ symmetric S-folds. Although some preliminary checks of non-perturbative stability against decay through standard brane-jet instabilities \cite{Bena:2020xxb} and nucleation of bubbles of nothing \cite{Witten:1981gj,Ooguri:2017njy} were performed in \cite{Giambrone:2021wsm}, a final answer to the question of the non-perturbative stability remains elusive. Also a full characterisation of the (possibly non-supersymmetric) conformal manifold of S-fold CFT's is still missing. Along these lines, an interesting puzzle regarding the compactness of the marginal deformation preserving $\,\mathcal{N}=2\,$ supersymmetry and connecting the $\,\mathcal{N}=4\,$ and $\,\mathcal{N}=2\,$ S-folds
has been stated in \cite{Bobev:2021yya} (see discussion below eq.(\ref{F5_solu_N2})) posing some challenges to the CFT Distance Conjecture \cite{Perlmutter:2020buo}. Finally, it would also be interesting to search for S-fold solutions in less supersymmetric setups. These are certainly very interesting open questions and research lines to be investigated in the near future.

\section*{Acknowledgements}

We are very grateful to Alfredo Giambrone, Emanuel Malek, Henning Samtleben and Mario Trigiante for collaboration in the related work \cite{Giambrone:2021wsm}. Especially to Mario Trigiante for collaboration also in \cite{Guarino:2020gfe}. AG is supported by the Spanish government grant PGC2018-096894-B-100. CS is supported by IISN-Belgium (convention 4.4503.15) and is a Research Fellow of the F.R.S.-FNRS (Belgium).

\bibliographystyle{JHEP}
\bibliography{references}

\end{document}